\documentclass{ws-procs9x6}

\begin{document}

\title{The Lamb Shift and Ultra High Energy Cosmic Rays}

\author{She-Sheng Xue}

\address{I.C.R.A.
and
Physics Department, University of Rome ``La Sapienza", I-00185 Rome,
Italy\\ 
E-mail: xue@icra.it}


\maketitle

\abstracts{
On the analogy with the Lamb shift, we study the vacuum effect that proton's electric field 
interacts with virtual particles in the vacuum. We find a possible quantum instability that 
triggered by an external force, proton's electric field interacting with virtual particles
spontaneously induces a quantum force that back reacts on the proton in the direction of 
the external trigger force. Such a quantum-induced force accelerates the proton runaway, 
by gaining the zero-point energy from the vacuum ($\sim 10^{-5}$ eV/cm). 
This effect possibly accounts for the mysterious 
origin and spectrum of ultra high-energy cosmic ray (UHECR) events above $10^{20}$eV, 
and explains the puzzle why the GZK cutoff is absent. The candidates of these events 
could be primary protons from the early Universe.}

\noindent{\it The effective Lagrangian for a proton.}\hskip0.3cm
The Lamb shift\cite{lamb} shows that the energy level $2S_{1\over2}$ of the hydrogen atom spectrum is 
shifted {\it upward} $+1008$MHz, compared with $2P_{1\over2}$. 
This implies that QED vacuum effects drain the zero-point energy to a hydrogen atom. 
We relate these vacuum effects to the origin of UHECR events.

Considering a proton interacting with virtual particles in the vacuum, 
we introduce (i) $\Psi$ and $A_\mu$ describing a proton field and its gauge potential;
(ii) $\psi_q$ and $A^q_\mu$ describing the quantum fields of virtual fermions and 
photons in the vacuum. 
To study this system, we start with a renormalized lagrangian 
density $L(x)$ with all necessary renormalization counterterms,
\begin{eqnarray}
\! L(x)&\!=\!& -\!{1\over4}(F^2+F^2_q)\!+\!\bar\Psi\big[i\gamma^\mu\partial_\mu \!
-\!m_p\!-\!e_p\gamma^\mu 
(A_\mu\!+\! A^q_\mu)\big]\Psi\nonumber\\
&\!+&\!\bar\psi_q\big[i\gamma^\mu\partial_\mu - m 
-e\gamma^\mu (A_\mu+ A^q_\mu)\big]\psi_q\!+\!({\rm c.t.}),
\label{l}
\end{eqnarray}
where $F$ and $F_q$ are classical and quantum electromagnetic field tensors, $e$ and $m$
($e_p$ and $m_p$) and  
are electron(proton) charge and mass. This is a complex interacting system, 
the classical fields $\Psi$ 
and $A_\mu$, quantum fields $\psi_q$ and $A^q_\mu$ are coupled together. 
To the first order, we obtain an effective interacting lagragian (details will be presented elsewhere):
\begin{eqnarray}
\! L_i^{\rm ext}(x)\!&\!=\!&\!-\!\bar e A_\nu^{\rm ext}\bar\Psi\gamma^\nu\Psi,\hskip0.3cm
\bar e\!=\!e_p{4\alpha^2\over3}\big(\ln{m\over\mu}\!-\!{7\over40}\big);\label{effi}\\
\!A_\nu^{\rm ext}(x)\!&\!=\!&\!\int\!{d^4x'\over m^2}{\rm tr}\left[S_F(x\!-\!x')\gamma_\mu S_F(x'\!-\!x)\gamma_\nu\right]A^\mu(x'),
\label{effa}\\
\!&\!=\!&\!{1\over 60\pi m^4}
\left( g_{\mu\nu}-{\partial_\mu \partial_\nu\over \Delta }\right)(\Delta)^2A^\mu(x).
\label{effak}
\end{eqnarray}
In Feynman's prescription of particles and antiparticles, Eq.(\ref{effa}) shows that a
pair of virtual fermion and antifermion ({\it virtual pair}) is created at one 
spacetime point $x$ and annihilates at another $x'$, behaving as an electric dipole $\vec P$
in its life-time. This {\it virtual pair} couples to the classical field $A^\mu(x')$ 
of the proton at $x$. As a result, an induced quantum dipole field $A_\nu^{\rm ext}(x)$ 
is created, attributed to {\it virtual pairs}, 
and this quantum field back interacts with the proton as an external field 
$A_\mu^{\rm ext}(\vec E^{\rm ext})$ . 

\vskip0.15cm
\noindent{\it Induced quantum force and instability.}\hskip0.3cm  
In the case of the proton at rest or traveling with a constant velocity, we might conclude 
$A_\mu^{\rm ext}(x)\equiv 0$ for its transversality and the Lorentz invariance: 
$A^\mu(x)={e_p\over 4\pi |\vec x|}g^{\mu\circ}$, being longitudinal in an 
instantaneous rest frame of the proton. However, we have to consider the quantum nature of
quantum-induced field $A_\mu^{\rm ext}(x)$. In the absence of an external field, the 
quantum-field fluctuations of {\it virtual pairs} and their 
dipole fields are entirely random-fluctuating of a time-scale $\delta\tau_q\sim 1/m$ 
in the spacetime, we do not expect any quantum-induced field of life-time $>\delta\tau_q$ 
in(at) any particular direction(point) 
of the spacetime. In the presence of longitudinal electric field of the proton, (i) the 
transverse quantum-field fluctuations of {\it virtual pairs} and their 
dipole fields $A_\mu^{\rm ext}(x)$ are entirely random-fluctuating of a time-scale $\delta\tau_q\sim 1/m$ in the spacetime; (ii) the longitudinal quantum-field fluctuations 
of {\it virtual pairs} and their dipole fields $A_\mu^{\rm ext}(x)$, although their life-time 
can be larger than $\delta\tau_q$, are entirely spherically symmetric and total dipole field $A_\mu^{\rm ext}(x)$ acting on the 
proton is zero. Thus, indeed we do not expect any induced quantum field $A_\mu^{\rm ext}(x)$ 
of life-time $>\delta\tau_q$, acting on the proton in a peculiar direction of the 
instantaneous rest frame. 
 
Nevertheless, in the instantaneous rest frame of the proton, we consider the case that 
an external trigger force $\vec F_{tri}$ 
accelerates the proton for a time interval $\Delta t_{tri}\gg \delta\tau_q$. As a result, 
proton's electric field $\vec E(x')$ gets a transverse component $\vec E_{\perp}(x')$, whose
distribution (both value and direction) is axial symmetric with respect to 
the direction of $\vec F_{tri}$, as given by the Lienard Wiechert field.  
This transverse field induces the transverse component $\vec P_{\perp}(x')$ 
of quantum electric dipoles of {\it virtual pairs} at $x'$, 
$\vec P_{\perp}(x')\sim \vec E_{\perp}(x')$. The spatial distribution of 
$\vec P_{\perp}(x')$ is the same as that of $\vec E_{\perp}(x')$. These quantum electric 
dipoles create electric dipole fields $\vec E^{\rm ext}_{\perp}(x)$ (\ref{effak}) 
back reacting on the proton at $x$. 
Summing over all contributions of quantum electric dipoles $\vec P_{\perp}(x')$ of 
{\it virtual pairs}, we find that the total $\vec E^{\rm ext}_{\perp}(x)$ acts on the 
proton in the same direction of $\vec F_{tri}$.
This total $\vec E^{\rm ext}_{\perp}(x)$ acting on the proton then plays the same role of 
$\vec F_{tri}$.
This implies that quantum-field fluctuations of {\it virtual pairs}, triggered by 
$\vec F_{tri}$, could cause a quantum runaway instability that the proton is accelerated 
further and further by a quantum force $\vec F_{q}=\bar e\vec E^{\rm ext}_{\perp}(x)$ 
even after the trigger force is off. Such a quantum instability can take place, provided 
quantum electric dipoles $\vec P_{\perp}(x')$ and their electric fields 
$\vec E^{\rm ext}_{\perp}(x)$ have a life-time $\gg\delta\tau_q$. 
For the Lorentz invariance and homogeneity of the vacuum state, the quantum-induced field
$\vec E^{\rm ext}_{\perp}$ or $F^{\rm ext}_{\mu\nu}$ must be constant. 

In the following, we adopt a semi-classical model to qualitatively estimate the value of 
such an induced quantum driving force in the instantaneous rest frame of the proton. 
{\it Virtual pairs} in an external field can be possibly considered as unstable 
excitations of bound states of virtual fermions and antifermions. We approximately estimate their binding energy, size and life-time. 
The energy scale of quantum-field fluctuations of {\it virtual pairs} must be much smaller 
than the electron mass $m$, otherwise real electrons and positrons would be created. We thus 
adopt a non-relativistic description for {\it virtual pairs}, whose size is about 
${2\over\alpha m}$, binding energy $\sim {\alpha^2 m\over 2}$. This indicates   
the size of electric dipoles $|d|\sim |x-x'|\sim {2\over\alpha m}$, and $\vec P=|e|\vec d$ 
in Eq.(\ref{effa}). 
The cross-section(probability) of the annihilation and creation of such a {\it virtual pair}
is about $\pi({\alpha\over m})^2$. The life-time of such a {\it virtual pair} is 
then $\delta\tau_p\sim {2\over\alpha^5 m}=6.2\cdot 10^{-11}$sec. 
This indicates the life-time of quantum electric dipole 
$\vec P_{\perp}$ and field $\vec E^{\rm ext}_{\perp}(x)$ is 
$\delta\tau_p\sim {2\over\alpha^5 m}$, which is much larger than 
$\delta\tau_q\sim {1\over m}$. Using $\vec P_{\perp}\cdot \vec E_{\perp}\lesssim
e^2/(4\pi |d|)$, we can estimate $|\vec E_{\perp}|\lesssim |e|/(4\pi |d|^2)$.  

The large wavelength modes $k$
of proton's gauge field $A_\mu$ are sensitive to the low-lying states of {\it virtual pairs}
of size $\sim {2\over\alpha m}$. This suggests $k\sim {\alpha m\over2}$ in 
Eq.(\ref{effak}) and the infrared cutoff $\mu\sim {\alpha m\over2}$ in Eq.(\ref{effi}).
The amplitudes of the induced quantum dipole fields $A_\mu^{\rm ext}$ and $\vec E^{\rm ext}$ 
are approximately given by,
\begin{equation}
A_\mu^{\rm ext}\!\simeq\!{\alpha^4\over 960\pi}A_\mu;\hskip0.3cm
\vec E^{\rm ext}_{\perp}\!\simeq\!{\alpha^4\over 960\pi}\vec E_{\perp}.
\label{effake}
\end{equation}
Summing over the angular distribution of {\it virtual pairs}, we obtain the 
spontaneously induced quantum 
force:
\begin{equation}
\vec F_{q}={\delta E\over\delta \vec x} =\bar e\vec E^{\rm ext}_{\perp}
\simeq  2.82\cdot 10^{-5}({\rm eV/cm}) \vec u,
\label{rate}
\end{equation}
and its direction $\vec u (|\vec u|=1)$ is kept in the same direction of proton's
acceleration. The life-time of this induced quantum 
force is $\delta\tau_p\sim {2\over\alpha^5 m}\gg \tau_q$, and it seems that the quantum 
instability ought to occurs. Eq.(\ref{rate}) holds for any charged particles and 
can be experimentally tested in a laboratory. The estimations and considerations 
are still very qualitative and speculative, need to be further improved and verified. 
It is highly deserved to have a quantitative computation of this quantum-induced force 
and instability.

\vskip0.15cm
\noindent{\it Vacuum energy gain and lost.}\hskip0.3cm
We turn to discuss Eq.(\ref{rate}) from the energetical point of view.
In the absence of any external field, the quantum-field fluctuations 
of {\it virtual pairs} are entirely random in the spacetime. This determines 
the maximum value of the zero-point energy. However, in the presence of a proton 
and its external field that couples to {\it virtual pairs}, the quantum-field 
fluctuations of {\it virtual pairs} 
are re-oriented towards the direction of the external field, so that the zero-point 
energy is reduced. The variation of the zero-point energies due to the longitudinal 
component of the external field dissipates back to the vacuum and the external field. 
While, the variation of the zero-point energies due to the transverse component of 
the external field drains to the proton as 
a recoiling effect. This recoiling effect is realized by an induced quantum field 
$A_\mu^{\rm ext}$ (\ref{effa}) back reacting on the proton.
The re-orientation of quantum-field fluctuations of {\it virtual pairs} 
towards external field's direction takes place during their life-time
$\delta\tau_p\sim 6.2\cdot 10^{-11}$sec.~(corresponding to $1.86$cm). The zero-point 
energy variation $\delta\epsilon\sim\alpha^5 m= 5.2\cdot 10^{-6}$eV, given 
by the Heisenberg uncertainty relationship, consistently with the rate (\ref{rate}) 
of the zero-point energy variation. 

We discuss a proton passing through the vacuum. After triggered, the proton 
driven by the quantum-induced force moves from one spacetime point to another, 
{\it virtual pairs} are involved in 
interacting with the transverse component of proton's electric field, more and more 
the zero-point energy drains into the proton. As a consequence, the constant 
quantum-induced force, which is rather analogous to the Casimir force, is built to 
accelerates the proton, as if the proton gets a continuous recoil from the vacuum and rolls 
down along a potential with a very small slop $\sim -10^{-5}$eV/cm. 
In this spontaneous process, the proton gains the zero-point energy and 
the vacuum reduces its zero-point energy in such a way that the whole interacting system 
of the vacuum and proton minimizes its interacting energy. This causes the energetically favourable instability and accelerating the proton runaway. 

However, any other trigger forces $\vec F'_{tri}$ acting on the proton can alter the 
direction of the quantum-induced force $\vec F_q$, since the later always keeps in 
the same direction of acceleration of the proton. Let us consider the following case:
a proton driven by $\vec F_q$ (\ref{rate}) moves in velocity $\vec v$ that is in 
the same direction of $\vec F_q$; a trigger force $\vec F'_{tri}$ acts on such 
a proton in the opposite direction of $\vec F_q$ and $|\vec F'_{tri}|>|\vec F_q|$. 
The direction of $\vec F_q$ is altered to the direction of $\vec F'_{tri}$. After the 
trigger force $\vec F'_{tri}$ is off, the direction of $\vec F_q$ turns out to be opposite
to the direction of proton's velocity $\vec v$. This causes the de-acceleration
of the proton, Eq.(\ref{rate}) is negative for energy-lost, indicating that the kinetic 
energy of the proton drains back to the zero-point energy of the vacuum. In general, this
happens for $\vec F_q\cdot \vec F'_{tri}<0$ and $|\vec F'_{tri}|>|\vec F_q|$. Trigger forces 
$\vec F_{tri}$ acting on protons are attributed to all real particles and fields, 
rather than virtual particles in the vacuum. With respect to a 
proton, these trigger forces are totally random in the spacetime. This indicates 
that in our Universe, some protons gain the zero-point energy from the vacuum, whereas 
others instead lose their kinetic energy to the vacuum, both directions are equally 
probable and none of them is preferential. Our Universe is not continuously heated 
up by gaining the zero-point energy of the vacuum.

To discuss the energy conservation in such an spontaneously induced process of the 
matter and vacuum interaction, we would like to first take the Casimir effect (force) as an
analogue. The Casimir force(vacuum) drives two separating plates moving closer and closer 
at the cost of the zero-point energy of the vacuum. On the other hand, any external 
force(matter) drives against the Casimir force to separate two plates moving further 
and further, and makes an energy-drain back to the vacuum. The induced quantum force 
$\vec F_q$ accelerates particles at the cost of the zero-point energy of the vacuum, and
de-accelerates particles at the cost of the kinetic energy of particles. Energy-drain goes
in bath directions, back and forth in between the matter and vacuum. The total energy of 
our Universe is conserved.

\vskip0.15cm
\noindent{\it UHECRs.}\hskip0.3cm
Based on the rate of energy-gain (\ref{rate}) and considering those primary protons that 
the energy-gain prevails in their traveling, we give a very preliminary discussion on UHECRs. 
With the present size of the Universe $\sim 10^{28}$cm, protons can possibly reach 
the energy more than $10^{21}$eV, if they travel a distance $D$ of $10^{27}-10^{28}$cm before reaching us. 
In such a scenario, {\it primary protons}, the candidates of UHECR events, 
could be originated from the astrophysical sources of large redshift 
$z$, like Quasars, or from the early Universe, and no particular arrival 
direction can be identified. 

The GZK cutoff does not apply to such a process of protons gaining 
energy bit by bit on their way to us. The reason is that protons, beyond $\sim 50$Mpc from us,
have an energy much smaller than the energy threshold $10^{20}$eV. This explains the absence of
the GZK cutoff in UHECR events. However, when protons near us reach the 
energy $10^{21}$eV, the GZK effect acts and average energy loss is about $(10^{-5} - 10^{-6})$eV/cm\cite{aloss}, which is roughly in the order of energy gain (\ref{rate}). 
This implies that ultra high energy protons would not have large possibilities 
to exceed the energy $10^{22}$eV. 
 
We set the origin of a spherical coordinate at the center of a primary proton's source, 
whose size is $R_\circ$, number-density $n_\circ$ and mean outgoing velocity 
$v_\circ$. The total flux out of the source is  $4\pi R_\circ^2 n_\circ v_\circ$.  
The Earth is located at $R$ distance away from the source. The total flux passing through 
the spherical surface $4\pi R^2$ is $4\pi R^2 n v$, where $n$ is the 
number-density of UHECR protons and $v$ the mean velocity. We have the conservation of 
total numbers of UHECR protons:
\begin{equation}
4\pi R^2n v= 4\pi R_\circ^2n_\circ v_\circ (1+z)^{-3},
\label{conf}
\end{equation}
where the factor $(1+z)^{-3}$ is due to the effect of expanding Universe. Thus we obtain 
the flux of UHECRs measured on the Earth,
\begin{equation}
\Phi=4\pi R_\circ^2 n_\circ v_\circ (1+z)^{-3}{1\over4\pi R^2}\sim {1\over R^2}.
\label{fluxe}
\end{equation}
Due to the distribution of intergalactic magnetic field and/or galactic wind etc, 
protons normally travel in a zigzag way with a mean-free path 
$\lambda_p$. The distance $D$ that protons travel is certainly larger than $R$. 
In one extreme case, protons travel to us in a straightforward line, $D=R$. 
While in another extreme case, protons travel in a way of random 
walk, $D={R\over\lambda_p}R$. This gives rise to the spectrum of UHECR flux 
observed on the Earth: 
\begin{equation}
\Phi(E)\sim {1\over R^2}\sim {1\over E^\gamma}\hskip0.5cm 1\le\gamma\le 2,
\label{spec}
\end{equation}
where $\gamma=2$ is for proton traveling in a straightforward line ($\lambda_p=R$) and 
$\gamma=1$ in random walk.

\end{document}